\UseRawInputEncoding
\documentclass[%
 reprint,
superscriptaddress,
 amsmath,amssymb,
 aps,
prb,
]{revtex4-2}

\usepackage{graphicx}% Include figure files
\usepackage{dcolumn}% Align table columns on decimal point
\usepackage{bm}% bold math
\usepackage{multirow}
\usepackage{amsmath}
\usepackage{ulem}
\usepackage{xcolor}
\usepackage{hyperref}% add hypertext capabilities
\newcommand{\BSCCO}{Bi$_2$Sr$_2$CaCu$_2$O$_{8+\delta}$}
\newcommand{\NaCCOC}{Ca$_{2-x}$Na$_x$CuCl$_2$O$_2$}

\begin{document}

\preprint{APS/123-QED}

\title{Periodic Atomic Displacements and Visualization of the Electron-Lattice Interaction in the Cuprate}

\author{Zengyi Du}
\affiliation{Condensed Matter Physics and Materials Science Department, Brookhaven National Laboratory, Upton, NY, 11973 USA}%

\author{Hui Li}
\affiliation{Condensed Matter Physics and Materials Science Department, Brookhaven National Laboratory, Upton, NY, 11973 USA}%
\affiliation{Department of Physics and Astronomy, Stony Brook University, Stony Brook, NY 11790 USA}%

\author{Genda Gu} 
\affiliation{Condensed Matter Physics and Materials Science Department, Brookhaven National Laboratory, Upton, NY, 11973 USA}%

\author{Ahbay N. Pasupathy} 
\affiliation{Condensed Matter Physics and Materials Science Department, Brookhaven National Laboratory, Upton, NY, 11973 USA}%
\affiliation{Department of Physics, Columbia University, New York, NY 10027 USA}%

\author{John M. Tranquada} 
\affiliation{Condensed Matter Physics and Materials Science Department, Brookhaven National Laboratory, Upton, NY, 11973 USA}%

\author{Kazuhiro Fujita} 
\email{kfujita@bnl.gov}
\affiliation{Condensed Matter Physics and Materials Science Department, Brookhaven National Laboratory, Upton, NY, 11973 USA}%

\date{\today}% It is always \today, today,
             %  but any date may be explicitly specified
\begin{abstract}
	Traditionally, X-ray scattering techniques have been used to detect the breaking of the structural symmetry of the lattice, which accompanies a periodic displacement of the atoms associated with charge density wave (CDW) formation in the cuprate pseudogap states. 
	Similarly, the Spectroscopic Imaging Scanning Tunneling Microscopy (SI-STM) has visualized the short-range CDW. 
	However, local coupling of electrons to the lattice in the form of a short-range CDW has been a challenge to visualize, thus a link between these measurements has been missing. 
	Here, we introduce a novel STM-based technique to visualize the local bond length variations obtained from topographic imaging with picometer precision. Application of this technique to the high-$T_c$ cuprate superconductor \BSCCO revealed a high-fidelity local lattice distortion of the BiO lattice as large as 2$\%$. 
	In addition, analysis of local breaking of rotational symmetry associated with the bond lengths reveals modulations around four-unit-cell periodicity in both $B_1$ and $E$ representations in the $C_{4v}$ group of the lattice, which coincides with the uni-directional \textit{d}-symmetry CDW ($d$CDW) previously identified within the CuO$_2$ planes, thus providing direct evidence of electron-lattice coupling in the pseudogap state and a link between the X-ray scattering and STM measurements. 
	Overall, our results suggest that the periodic lattice displacements in E representations correspond to a locally-frozen version of the soft phonons identified by the X-ray scattering measurements, and a fluctuation of the bond length is reflected by the fluctuation of the dCDW formation near the quantum critical point.  
\begin{description}
		
%		\item[DOI]
		
		\item[Subject Areas]
		Condensed Matter Physics, Strongly Corrected Materials, Superconductivity
	\end{description}
\end{abstract}

%\keywords{Suggested keywords}%Use showkeys class option if keyword
                              %display desired
\maketitle

%\tableofcontents

\section{\label{sec:introduction}Introduction\protect\\}
The CuO$_2$ plane is a common crystallographic ingredient among different cuprate families and is the main playground of the electronic system. 
It determines both the electronic and magnetic characteristic properties, and exhibits instabilities towards states such as antiferromagnetism, $d$-wave superconductivity [\onlinecite{PairingSymmetry}], CDW [\onlinecite{CDW}], electronic nematicity [\onlinecite{ES}, \onlinecite{NSTM}, \onlinecite{NematicityTransport}], and pair density wave (PDW) [\onlinecite{PDWReview}, \onlinecite{ReviewOurs}]. 
Interactions among these states as well as their relationships to the bosonic excitations such as the antiferromagnetic spin fluctuation and the phonon excitations underlie the grand challenge of understanding the mechanism of the cuprate high temperature superconductivity.  

\begin{figure*}[htbp]
\centerline{\includegraphics[scale=.34]{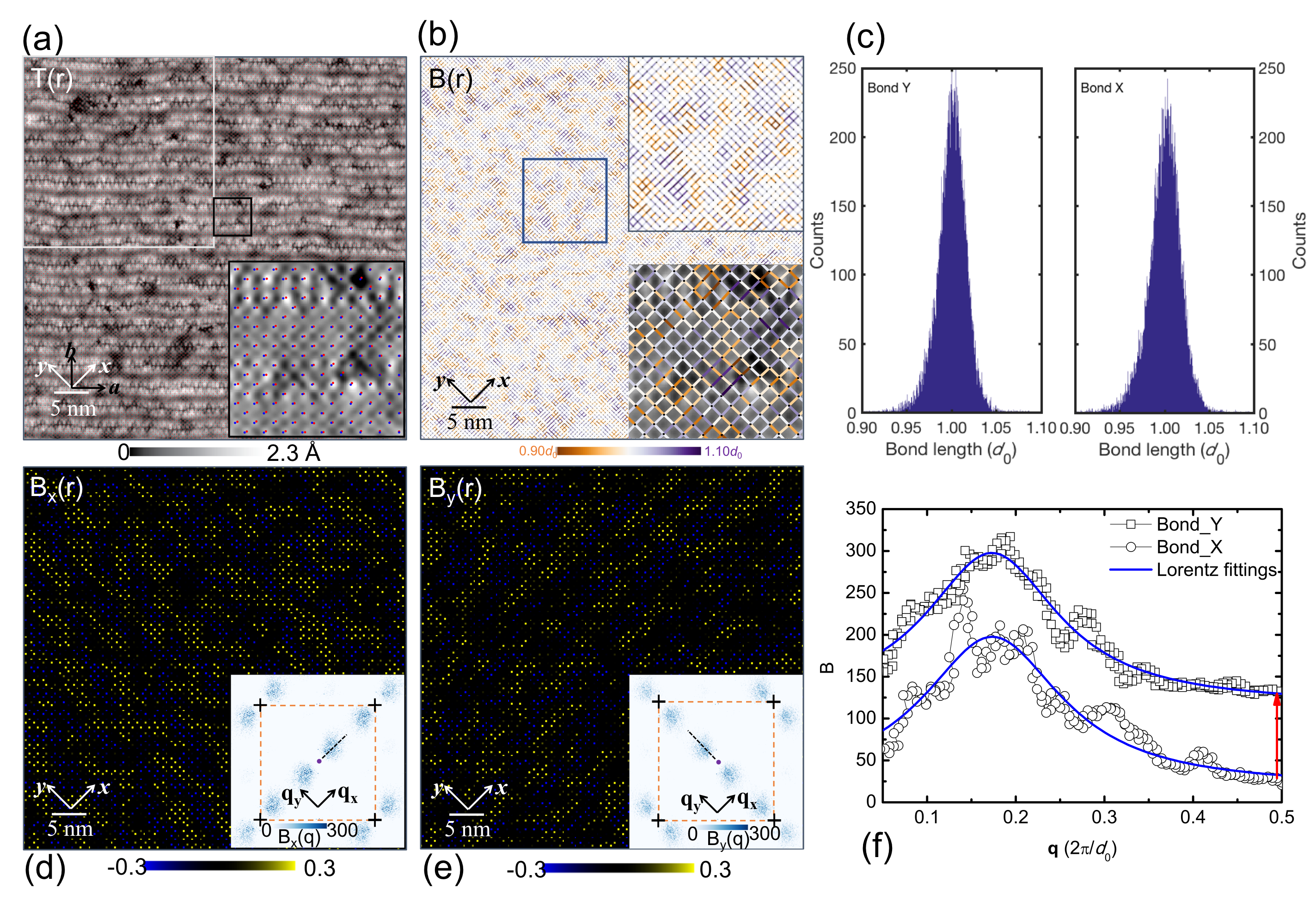}}
\caption{Spatial variation of the bond length
(a), Topograph of the \BSCCO taken within 50 nm $\times$ 50 nm FOV. 
\textcolor{black}{Inset shows a zoomed in topograph from the small rectangle region near the center of the FOV. 
Bi locations are marked up by the red dots, while perfect lattice locations are marked up by the blue dots. }
(b), Bond length map, in which adjacent Bi atoms are connected by the straight lines. Line colors represent a bond length in the unit of $d_0$ ($\sim$3.83\AA).
\textcolor{black}{Inset shows the same topo as in the inset in (a) superimposed with the bond length map. }
(c), Histograms of the bond length for both $x$ and $y$ directions. 
(d)[(e)], A sublattice representation of the bond length along $x$ ($y$) direction. 
A value of the bond length is assigned at O$_x$ (O$_y$) position of the CuO$_2$ plane. An inset is a magnitude fourier transform of $B_x$(\textbf{r}) [$B_y$(\textbf{r})]. 
(f), Line profiles of the fourier transform in (d) and (e) insets extracted from the black dashed lines along the Cu-O-Cu bond directions. 
The black crosses are locations of the Bragg peaks. 
}
\label{fig1}
\end{figure*}

One of the key phenomena commonly associated with the cuprate pseudogap state is CDW order, for which a static modulation of the charge appears below or near the pseudogap opening temperature $T^*$ [\onlinecite{YBCOxRS}, \onlinecite{CDWinBSCCO}]. 
Such a CDW has been ubiquitously observed in different cuprate families, and the associated period is 3-4 unit cells, depending on the material[\onlinecite{CDW}]. 
SI-STM measurements on \BSCCO and \NaCCOC has provided evidence for a unidirectional electronic modulation with $d$-wave symmetry ($d$CDW), in which the charge modulations primarily occur at in-plane Oxygen sites, O$_x$ and O$_y$, in a fashion such that the changes on these sites are out of phase by $\pi$[\onlinecite{LFCorrection}]. 
In contrast, X-ray diffraction on different cuprates provides evidence of periodic modulations of the lattice, from which CDW (or charge-stripe) order is inferred[\onlinecite{CDWinYBCO}, \onlinecite{StripeinLBCO}], while resonant soft-X-ray scattering provides evidence of charge modulations on specific atomic sites [\onlinecite{ModulationinLBCO}, \onlinecite{RIXSEnergy}].
Furthermore, neutron scattering, and resonant inelastic X-ray scattering (RIXS) measurements show a softening of bond-stretching phonons at the wave vector corresponding to charge order [\onlinecite{LOPhononLBCO}, \onlinecite{EPCTranquada}, \onlinecite{CDWaroundQCP}]. 
An open question is whether it is possible to detect in real space the lattice modulation associated with the dCDW in \BSCCO.
Here we demonstrate a new approach to analyzing local bond-length modulations detected by STM topography and show how they correlate with the electronic modulations measured simultaneously by SI-STM. 

Fundamentally, the CuO$_2$ plane is formed by a square lattice of Cu, with Oxygens O$_x$ and O$_y$ bridging Cu atoms along orthogonal directions. 
In \BSCCO there is an orthorhombic distortion of the lattice along axes at 45° to the Cu-O-Cu bond directions, due to a long-period modulation within the BiO bilayers. 
The SI-STM measurements are performed after cleaving the sample, which leaves a BiO layer at the surface. 
The long-period modulation of the BiO layers is well characterized and can be mathematically removed from the topographic image, leaving on average a square lattice of Bi atoms whose positions are expected to correlate with the Cu atoms below them.

\section{\label{sec:bondlength}Symmetry of the CuO$_2$ square lattice\protect\\}
 The CuO$_2$ square lattice is group-theoretically classified as a $C_{4v}$ group consisting of 8 symmetry operations, $\{I, C_4, C_2, C_{4}^{3}, \sigma_{x}, \sigma_{y}, \sigma_d, \sigma_{d'}\}$, in which $I$ describes a unitary operation, $C_n$ describes an $n$-fold rotation, and  $\sigma_i$ describes a mirror operation along the $i$-plane [$d$ ($d'$) is a mirror plane 45° rotated from $x(y)$-axis].
 Irreducible representations of the  $C_{4v}$ group are often denoted by $A_1$, $A_2$, $B_1$, $B_2$, and $E$ [\onlinecite{GroupTheory}], in which $A$ represents a quantity that is invariant under $C_4$ rotation and $B$ represents one that changes the sign under the $C_4$ rotation. 
 $E$ is a two-dimensional irreducible representation. Ideally, physical quantities within the CuO$_2$ plane are invariant under these operations and are represented by the irreducible representations. 
 However, in a real material, there are impurities, missing atoms, and electronic instabilities coupling to the lattice that would break the underlying symmetry of the lattice. 
 By evaluating symmetries for each irreducible representation with respect to bond lengths about each atomic site, one can test for and characterize the spatial correlations of local lattice distortions. 

Before such an analysis can be attempted, we must first correct for an experimental artifact. 
The piezo scanner that controls the STM tip scanning along $x$ and $y$ directions at a pico-meter precision shows non-linear drifts such that atomic modulations observed in a topography $T$(\textbf{r}) are extrinsically phase disordered. 
In other words, an observed image of the atoms is not simply described by a sinusoidal wave, but by a combination of cosine and sine waves due to the extrinsic phase disorder. 
In fact, if the Fourier transform of a raw $T$(\textbf{r}) is evaluated, the Bragg peaks that correspond to periodicities of the atoms show finite intensities in the imaginary part (the sine component of the modulation) because of the tip drifts. 
Similarly, electronic structure images that are simultaneously taken with the $T$(\textbf{r}), such as current and d\textit{I}/d\textit{V} etc, are phase disordered in the same way.

To overcome this systematic error in the measurement, the Lawler-Fujita algorithm [\onlinecite{NSTM}]  has been widely used to remove the effect of the tip drift from images [\onlinecite{LFillya}, \onlinecite{HanaguriVortex}, \onlinecite{SROQPI}]. 
This algorithm derives a displacement vector field $\textbf{u}(\textbf{r})$ (Fig. \ref{fig1s} in the Appendix section) that describes local phase shifts of a particular modulation (in the present case, atomic modulations) at a coarse-graining length. 
For the piezo scanner drift, a coarse graining length scale is typically much longer than the atomic spacing $d_0$ (3.83 \AA) of the CuO$_2$ unit-cell since relaxation of the piezo scanner occurs very slowly. 
Here, we use $\textbf{u}(\textbf{r})$ to remove the piezo scanner drift rendering the $T(\textbf{r})$ and any simultaneously taken images to be lattice periodic \textcolor{black}{ on average over the } 30 \AA coarse graining length (Appendix \ref{Appendix:DetermineAtoms}: Determination of the atomic sites). 
To \textcolor{black}{ determine the residual (intrinsic) atomic displacements } and extract the bond length\textcolor{black}{s} between adjacent Bi atoms, we apply the Lawler-Fujita algorithm once again to obtain the displacement vector field $\textbf{u}'(\textbf{r})$, but now with 6 Å for the coarse graining length.
Now, $\textbf{u}'(\textbf{r})$ represents how much individual atoms are displaced from the perfect lattice configuration, allowing us to locate Bi positions and possible intrinsic local lattice distortions (Appendix \ref{Appendix:DetermineAtoms}: Determination of the atomic sites). 
\textcolor{black}{While the Cu atoms within CuO$_{2}$ plane are not directly identified as the STM scans over the BiO surface, as we shall show in Fig. \ref{fig4}, electronic structures from the CuO$_{2}$ layer spatially correlate with structural modulations of the atomic displacements detected at the BiO layer, suggesting that displacements of the Bi atoms reflect those for the Cu atoms within the CuO$_{2}$. }

\section{\label{sec:results}Results\protect\\}

We performed SI-STM measurements at 11 K on a single crystal of \BSCCO that is near optimal doping, with a superconducting transition temperature $T_c$ equal to 91 K (Appendix \ref{Appendix:STM}: SI-STM Measurements). 
Figure \ref{fig1}(a) shows a typical drift-corrected topography, $T(\textbf{r})$, of the \BSCCO sample in a 50nm $\times$ 50nm field of view (FOV). 
The atoms seen in the  $T$(\textbf{r}) are Bi, and the Cu atoms that form the two-dimensional CuO$_2$ plane where all the cuprate phenomenology occurs, are located ~6 Å below these Bi atoms. 
It should be noted that the crystal supermodulation displaces the atoms in the bulk at \textbf{q}=$\textbf{Q}_{SM}$ along the b direction. 
Such an effect is apparently stronger in BiO than in the CuO$_2$ planes[\onlinecite{4DIncomm}], masking intrinsic lattice displacements[\onlinecite{2D3DIncomm}]. 
It is also found that the amplitude of the modulation of the Bi and Cu atoms is, at most, 2$\%$ of the Cu-O-Cu bond length $d_0\sim$3.83 Å [\onlinecite{Zn}]. 
SI-STM visualizes both the spatial amplitude and phase of the modulation such that the effect of the supermodulation can be effectively removed by Fourier filtration (Fig. \ref{fig2s} of the Appendix section), after which we anticipate that the Cu atom positions can be precisely inferred from the Bi positions. 
Similarly, the symmetry of the lattice in the  CuO$_2$ plane can be obtained by studying the symmetry of the BiO plane that is observed in the $T(\textbf{r})$. 
Hence, local lattice distortions of the CuO$_2$ plane can be accessed by studying the corrected topography, $T(\textbf{r})$.

Now that we know where all the Bi atoms are in $T(\textbf{r})$, after the filtration of the supermodulation, we show in Fig. \ref{fig1}(b) a bond length map, in which the bond length is represented by a colored bar connecting nearest-neighbor Bi atoms. 
Small black dots represent the locations of the Bi atoms. 
It is obvious that the Bi atoms are disordered, exhibiting local displacements from the perfectly periodic positions. 
Figure \ref{fig1} (c) shows histograms of the bond lengths for both $x$ (right) and $y$ (left) directions. 
Both bond lengths show virtually identical distributions; peak positions (or averages) are located at \textbf{d}$_0$ and the full width at half maximum is about 2$\%$ of \textbf{d}$_0$ for both directions. 
Such disorder is somewhat larger than the estimate from neutron scattering measurements for a stripe-ordered system [\onlinecite{NeutronLSCO}]. 

\begin{figure}[htbp]
\centerline{\includegraphics[width=\columnwidth]{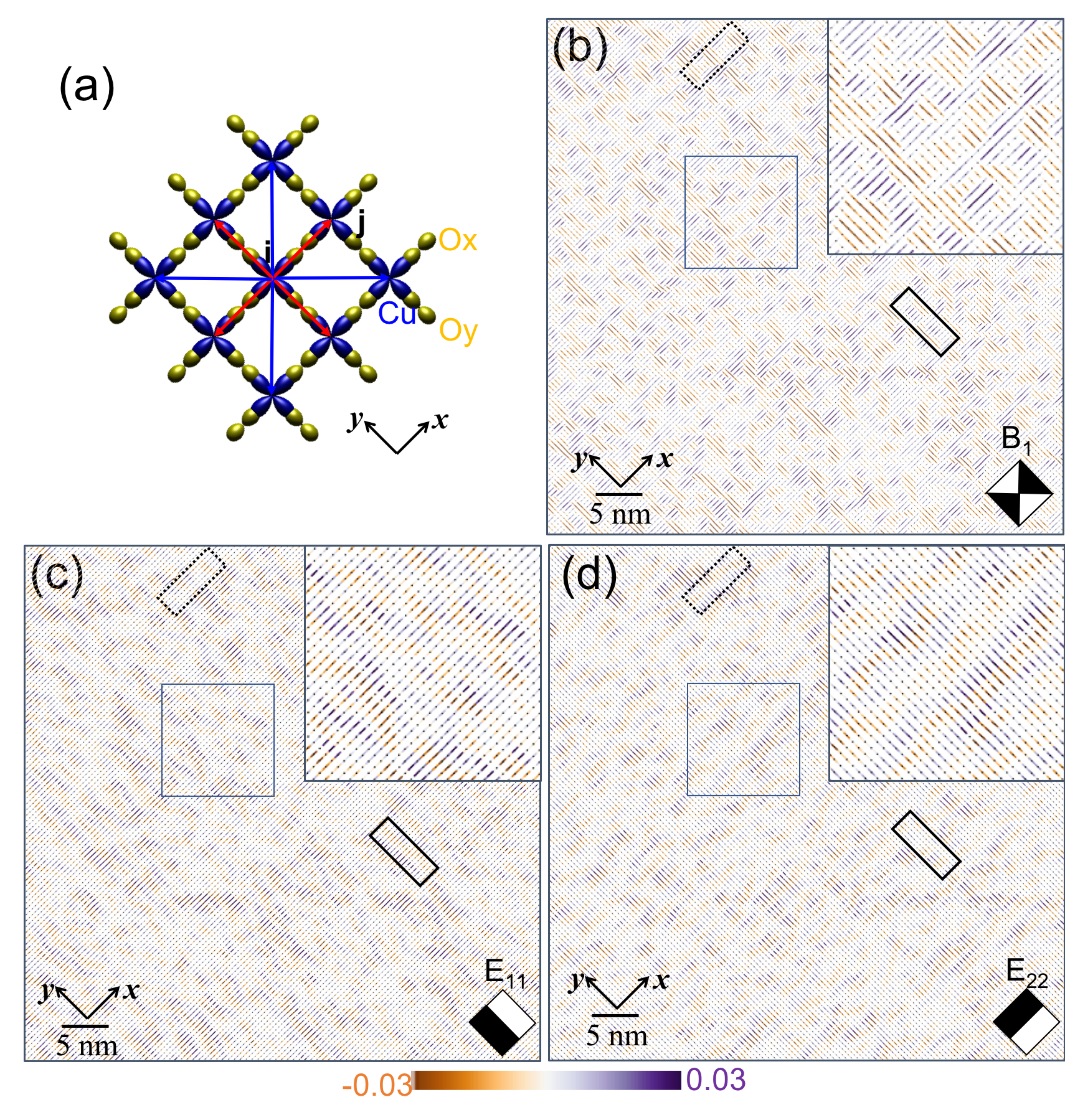}}
\caption{Symmetry representations of the in-plane bond length
(a), A schematic image of the CuO$_2$ plane, in which Cu 3$d_{x^2-y^2}$ orbitals and O $2p$ orbitals are represented in blue and yellow, respectively. 
The red and blue arrows are vectors connecting the nearest and the next nearest neighbor Cu sites. 
(b), $B_1$ representation of the bond length obtained from the topography shown in Fig.\ref{fig1}(a).
(c), (1,1) component of the $E$ representation, $E_{11}$, of the bond length obtained from the topography shown in Fig. \ref{fig1}(a). 
(d), (2,2) component of the $E$ representation, $E_{22}$, of the bond length obtained from the topography shown in Fig.\ref{fig1}(a). 
The solid (dotted) black rectangles in (b), (c), and (d) highlight locations of the canonical \textit{d}-symmetry CDW identified in the electronic structure for $x(y)$\textit{} direction\textcolor{black}{(Appendix Fig.\ref{fig4s})}.}
\label{fig2}
\end{figure}

Next, we separate the bond lengths according to orientation, associating the bond length between neighboring Bi atoms at \textbf{r}$_i$ and \textbf{r}$_j$ with the bond center \textbf{R}$_{ij}$=($\textbf{r}_i$+$\textbf{r}_j$)⁄2.
Note that, geometrically, \textbf{R}$_{ij}$ is approximately the location of the Oxygen sites, O$_x$ and O$_y$ in the CuO$_2$ plane. 
Figure \ref{fig1}(d) [(e)] is an intensity map of the bond length along the x-direction (y-direction), $B_x$(\textbf{r}) [$B_y$(\textbf{r})].
It is obvious that the bond lengths for both the x and y directions show strong modulations along the Cu-O-Cu directions.
The inset in Fig. \ref{fig1}(d) [(e)] is a Fourier transform of the $B_x$(\textbf{r}) $[$B$_y($\textbf{r})$]$, which exhibits a strong peak around \textbf{Q}$_B\sim$($\pm$0.2,0) [(0, $\pm$0.2)] similar to those observed for the electronic density wave in the pseudogap states [\onlinecite{dCDW}]. 
The same peaks also appear around the Bragg peaks (marked as “+”), due to the convolution of the lattice and modulations in real space [\onlinecite{LFCorrection}]. 
Line profiles shown in Fig. \ref{fig1}(f) reveal fine structures of these Fourier peaks for both the x and y directions, while the general shape of each line profile can be fitted by a Lorentzian. 
The fine structure suggests that there would be a domain structure of the bond length modulations having a discommensuration between the domains [\onlinecite{Commensureate4a0}]. 
These observations obviously show that the translational symmetry of the lattice is broken, which implies an associated modulation of electronic states.

\begin{figure*}[htbp]
\centerline{\includegraphics[scale=.36]{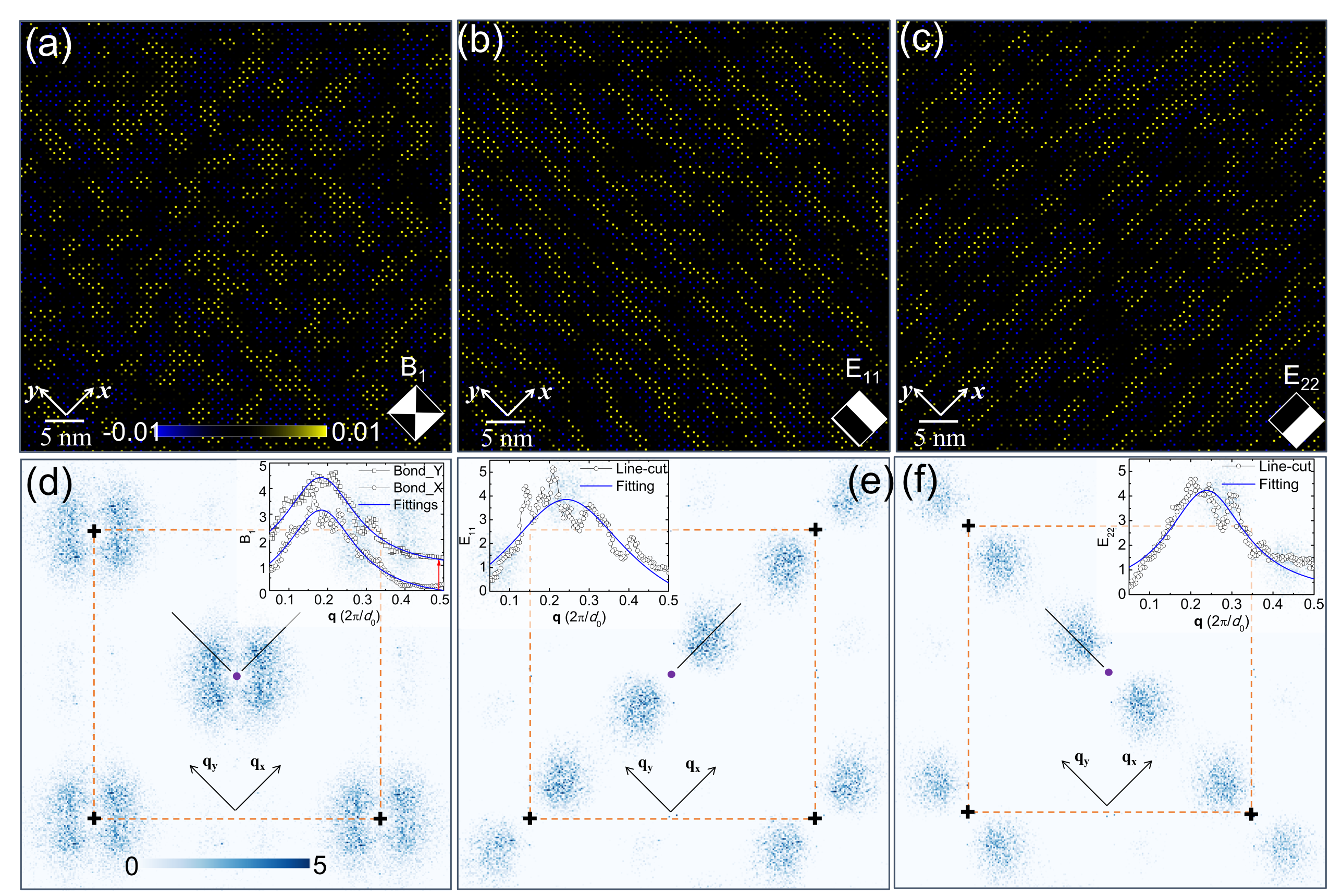}}
\caption{Spatial modulations of the symmetry 
(a), A sublattice representation of Fig.\ref{fig2}(b), $B_1$(\textbf{r}), in which spatial modulations are clearly resolved for both the x and y Cu-O-Cu bond directions. 
(b), A sublattice representation of Fig.\ref{fig2}(c), $E_{11}$(\textbf{r}), in which unidirectional modulations are clearly resolved for the x Cu-O-Cu direction. 
(c), A sublattice representation of Fig.\ref{fig2}(d), $E_{22}($\textbf{r}), in which unidirectional modulations are clearly resolved for the y Cu-O-Cu direction. 
(d), A magnitude Fourier transform of (a), exhibiting peaks along both the x and y Cu-O-Cu bond directions. 
An inset shows a line profile extracted from the black lines. The inset shows line profiles extracted from the black lines. 
(e), A magnitude Fourier transform of (b), exhibiting peaks along the $x$ Cu-O-Cu bond direction. An inset shows a line profile extracted from the black line. 
Data points (open circles) are fitted by the Lorentzian and a peak position is found at  \textbf{q}$_x$ = (0.242 $\pm$ 0.03) 2$\pi$/\textit{d}$_0$. 
(f), A magnitude Fourier transform of (c), exhibiting peaks along both the y Cu-O-Cu bond direction. 
An inset shows a line profile extracted from the black line. 
Data points (open circles) are fitted by a Lorentzian and a peak position is \textbf{q}$_y$ = (0.238 $\pm$0.03) 2$\pi/d_{0}$. The black crosses are locations of the Bragg peaks.
}
\label{fig3}
\end{figure*}

Now, we introduce a symmetry decomposition of the local lattice to analyze possible local breaking of rotational symmetry in the lattice structure.
We define a function  
\begin{equation}
S^\Gamma\left(\textbf{r}_{i}\right) = \frac{\sum_js^\Gamma_{ij}\vert \textbf{r}_i-\textbf{r}_j \vert}{\frac{1}{N}\sum_j\vert \textbf{r}_i-\textbf{r}_j \vert}.
\label{1}
\end{equation}
where $\Gamma$ is an irreducible  representation of the $C_{4v}$ group ($A_1$,$A_2$,$B_1$,$B_2$,$E$),  $\textbf{r}_i$ and $\textbf{r}_j$ are positions of the Bi atoms at $i$ and $j$ sites, respectively, and $s_{ij}^\Gamma$ is a symmetry dependent structure factor. 
A summation is taken over the nearest-neighbor sites for $\Gamma=A_1$,$B_1$,$B_2$,$E$, so that $\vert \textbf{r}_i-\textbf{r}_j\vert$ is on the scale of $d_0$ and the denominator in eq.(\ref{1}) is a local average of the bond lengths. 
(For $\Gamma=A_2$, the summation is taken over the next-nearest-neighbor sites blue arrows). 
For example, when we consider the $A_1$ representation of the local bonds,$s_{ij}^{A_1}$=1 since the $A_1$ representation shows no sign change by the symmetry operations in $C_{4v}$ group (Appendix \ref{C4v}: Representation of the $C_{4v}$ group and bond length irreducible representations and Table \ref{table1}). 
Then the summation is taken over the nearest-neighbor sites [red arrows in Fig. \ref{fig2}(a)] with the weight $s_{ij}^{A_1}$. 
To visualize the irreducible representations of the bonds obtained using eq. (\ref{1}), we draw a line over the atomic site $i$ with a color that represents the intensity obtained by eq. (\ref{1}). 
For example, for the $B_1$ representation [Fig. \ref{fig2}(b)], we draw a line along the direction of the expanded bonds[\onlinecite{PseudogapMott}].

\begin{figure}[htbp]
\centerline{\includegraphics[width=\columnwidth]{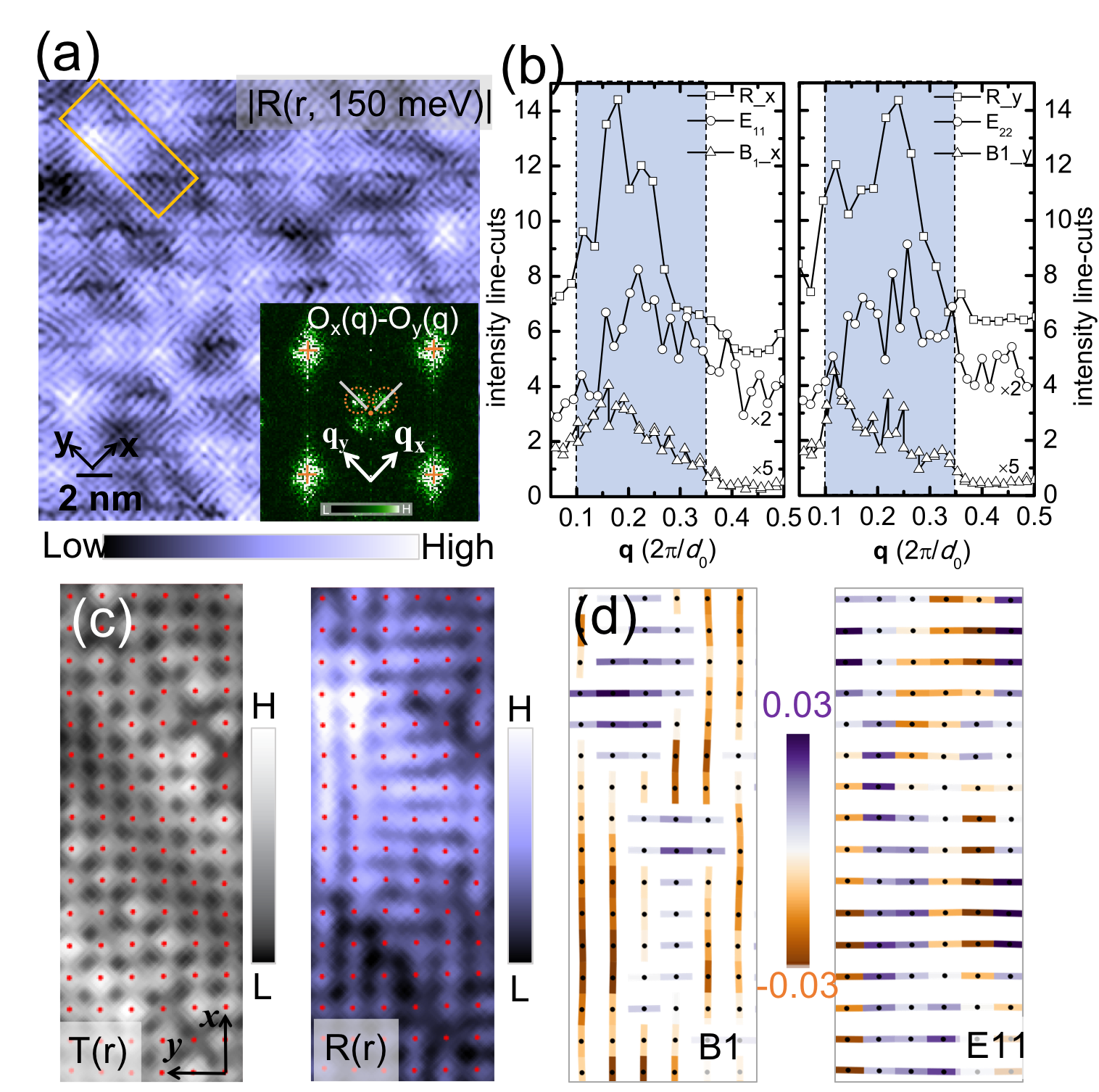}}
\caption{Direct visualization of the electron-lattice interaction
(a), $R$(\textbf{r},150mV) within 20nm x 20nm FOV exhibiting the $d$-symmetry CDW, in which electronic degree of freedoms reside primarily on the O$_x$ and O$_y$ modulating with out-of-phase by $\pi$. 
The inset shows the magnitude of the phase-resolved fourier transform of $R$(\textbf{r},150mV) in $d$-symmetry channel, exhibiting \textbf{Q}=($\pm$1/4,0); (0, $\pm$1/4) peaks as encircled by the orange dashed lines. 
The orange crosses mark locations of the Bragg peaks. 
(b), Line profiles extracted from the inset in A along grey lines, and line profiles along the same line trajectories in $B_1$(\textbf{r}), $E_{11}$(\textbf{r}) and $E_{22}$((\textbf{r}) for both $x$ (left) and $y$ (right) Cu-O-Cu bond directions. 
(c), $T$(\textbf{r}) and $R$(\textbf{r},150mV) extracted from a region in the solid orange rectangle in a, where a canonical $d$-symmetry CDW is observed. 
(d), $B_1$ and $E_{11}$ maps extracted from the same region of $B_1$(\textbf{q}) and $E_{11}$(\textbf{q}) obtained from the same FOV shown in (a).
}
\label{fig4}
\end{figure}

The $B_1$ representation is of particular interest, as it corresponds to a bond stretching quadrupolar mode [\onlinecite{LOPhononLBCO},\onlinecite{Tsuei}]. 
In Figure \ref{fig2}(b), we show the $B_1$ representation of the bond lengths, in the same FOV as Fig. \ref{fig1}; it exhibits a pattern in which blue and orange bonds alternate at a period of ~four-unit cells. 
This can be viewed as a bond “nematicity”. 
Moreover, a careful examination of the pattern within regions highlighted by the black rectangles in Fig. \ref{fig2}(b) leads us to identify a modulation similar to the unidirectional dCDW that can be seen in the electronic states [Fig. \ref{fig4}(a)]. 
In fact, the $B_1$ representation is equivalent to $d$-symmetry suggesting that $d$CDW couples to the lattice and locally alters the bond length in a $d$-wave fashion [\onlinecite{Tsuei}]. 

Figures \ref{fig2}(c) and (d) show the (1, 1) and (2, 2) components of the E representation, E$_{11}$ and $E_{22}$, respectively, as given by a 2 x 2 matrix representation. 
$S^{E_{11}}$ ($\textbf{r}_i$) [$S^{E_{22}}$ ($\textbf{r}_i$)] is simply the difference of the bond lengths between nearest neighbor sites of $\textbf{r}_i$ only along the $x$ ($y$) direction, characterizing a unidirectional stretching or shrinking of the bond length. 
It is intriguing that both the $E_{11}$ and $E_{22}$ representations show spatial modulations of the bond stretching. 
Similar to the $B_1$ representation shown in Fig. \ref{fig2}(b), a canonical “bond-stretching half-breathing mode” can be found in regions highlighted within the black rectangles again where unidirectional $d$CDW can be clearly observed. 
Other representations such as $A_1$, $A_2$ and $B_2$ are shown in the Fig. \ref{fig4s} of Appendix section and discussed in the Appendix \ref{C4v}: Representation of the $C_{4v}$ group and bond length irreducible representations.

To analyze the modulated structures of each representation of the bond length, shown in Fig.\ref{fig2}, we replot $S^{\Gamma}$($\textbf{r}_i$) for $\Gamma=B_1$, $E_{11}$, and $E_{22}$ in Fig.\ref{fig3}(a), (b), and (c), respectively, now with amplitude indicated by colored circles. 
The corresponding Fourier-transform magnitudes are presented in Fig. \ref{fig3}(d), (e), and (f), respectively. 
For the case of $B_1$ symmetry, the magnitude of the Fourier transform shown in Fig. \ref{fig3}(d) exhibits strong peaks near \textbf{Q}$\sim$($\pm$1/4, 0), (0, $\pm$1/4). 
Note that there are similar peaks surrounding the Bragg peaks marked by the black cross. 
Similarly, the magnitude Fourier transforms for $E_{11}$ and $E_{22}$, shown respectively in Figs. \ref{fig3}(e) and (f), exhibit strong peaks near the same wavevectors, but only along the \textbf{Q}$_x$ or \textbf{Q}$_y$ direction.
 
In Fig. \ref{fig4}(a), we show $R$(\textbf{r},150mV)$=\frac{\textit{I}(\textbf{r},+150\textrm{mV})}{\textit{I}(\textbf{r},-150\textrm{mV})}$, the ratio of the tunneling current measured at opposite bias voltages, which has been demonstrated to represent the cuprate dCDW without ambiguity, since the ratio can effectively mitigate a systematic error in the measurement related to the junction formation [\onlinecite{Chapter3}]. 
As shown in Fig.\ref{fig4}(a), unidirectional density wave modulations along both $x$ and $y$ directions are clearly visualized, and the inset shows the phase-resolved sub-lattice Fourier transform in a $d$-symmetry channel[\onlinecite{LFCorrection}], clearly showing peaks near \textbf{q}$\sim$($\pm$1/4,0);(0, $\pm$1/4). 
This confirms that the electronic degrees of freedom at the O$_x$ and O$_y$ sites are out of phase by $\pi$, indicating that the short-range density wave holds \textit{d}-symmetry even at $p\sim0.17$. 
In Fig.\ref{fig4}(b), we show a comparison of the line profiles from $R$(\textbf{q},150mV), $S^{B_1}$(\textbf{q}),$S^{E_{11}}$(\textbf{q}), and \textit{S}$^{E_{22}}$(\textbf{q}) for both $x$ (left) and $y$ (right) directions, where they all show peaks near $\vert$\textbf{q}$\vert\sim0.25$. 
Note that peak positions of the line profiles for $R$(\textbf{q},150mV) and $S^{B_1}$(\textbf{q}) are not peaked at exactly $\vert$\textbf{q}$\vert=0.25$; this is most likely due to the discommensuration of the $d$CDW domains[\onlinecite{Commensureate4a0}]. 
\textcolor{black}{A precise estimation of the wavevectors for $B_1$, $E_{11}$ and $E_{22}$ and a study of their amplitudes and phases in relationship to the electronic structure is an intriguing direction for the future studies.}

\section{\label{sec:discussion}Discussion\protect\\}
Figures \ref{fig4}(c) and (d) summarize the situation occurring in both the electronic and lattice structures within a canonical dCDW that is extracted from the black rectangle in Fig. \ref{fig2}. 
The $d$CDW breaks both the rotational and the translational symmetry as shown in Fig. \ref{fig4}(c) (right) together with the simultaneous topography (left), in which local atomic displacements are clearly seen (red dots are the locations of atoms). 
Our observations indicate that the unidirectionality of the $d$CDW couples to the lattice, in which the bond length is periodically stretched and shrunk as revealed in the $E$ representation (half-breathing mode) [Fig. \ref{fig4}(d) right], having a relatively shorter bond length at the bond center of the $d$CDW [\onlinecite{PseudogapMott}]. 
Simultaneously, a modulated intra-unit-cell rotational symmetry breaking of a $d$-symmetry electronic degree of freedom couples to the lattice, displacing the atoms as revealed in the $B_1$ representation (quadrupolar mode) [Fig.\ref{fig4}(d) left]. 

From these results, one notices that the bond length modulations in the $B_1$ and $E$ ($E_{11}$, $E_{22}$) representations show a strong correspondence to the electronic properties of the $d$CDW seen in the pseudogap states. 
First, the $d$CDW is highly unidirectional, breaking both rotational and translational symmetries[\onlinecite{LFCorrection},\onlinecite{dCDW}]. 
Second, it is highly heterogeneous[\onlinecite{dCDW}]. 
Third, it is predominantly $d$-symmetry[\onlinecite{LFCorrection},\onlinecite{dCDW}]. Fourth, the periodicity of the $d$CDW modulation is close to four-unit cells[\onlinecite{LFCorrection},\onlinecite{Symmetry}]. 
Thus, the modulations in  $S^{B_1}$(\textbf{q}),$S^{E_{11}}$(\textbf{q}),and $S^{E_{22}}$(\textbf{q}) together with the strong correspondence to $R$(\textbf{r},150mV) indicate that there are significant electron-lattice interactions, reflecting the observation that the atoms are displaced in the same symmetrical fashion as those for the density wave in the electronic structure at the pseudogap energy scale. 
The presence of such coupling indicates that the modulations seen in $R$(\textbf{r},150mV) indeed carry a charge degree of freedom, that of a charge density wave with $d$-symmetry ($d$CDW), since an electron-lattice interaction requires a local charge density. 

In addition, it is natural to speculate that the periodic lattice displacements visualized in $E$ representations would be linked to the phonon dynamics reported in recent resonant X-ray measurement[\onlinecite{CDWaroundQCP}], in which the interplay of the bond-stretching phonons and charge order fluctuations are increased with lowering temperature below $T_c$. 
The wavevector at which the phonon anomalies occur is quite similar to the one observed in the $E$ representations, suggesting that the periodic lattice displacements in $E$ representations correspond to a locally-frozen version of the soft phonons and consistent with the lock-in mechanism of the $d$CDW by the electron-phonon coupling [\onlinecite{Lockin}] for our \BSCCO  sample. 
The lack of spatial homogeneity of the bond-length modulations, as shown in Fig. \ref{fig1}(b), can be linked to the quantum fluctuations of the density waves inside the superconducting dome that are associated with the corresponding quantum critical point located just beyond the optimal doping [\onlinecite{CDWaroundQCP},\onlinecite{Kazu2014}]. 

In this study, we have introduced a group theoretic symmetry decomposition algorithm to analyze the symmetry of a lattice as well as the local displacements of it in \BSCCO.
First, we visualized the heterogeneity of the in-plane bond lengths, for which the fluctuation is about 2$\%$ of the unit-cell distance. 
Second, the bond lengths themselves are spatially modulating at ~four-unit-cell periodicity. 
Third, $B_1$, $E_{11}$, and $E_{22}$ representations of the bond length are modulating at similar periodicity. 
Fourth, a strong correspondence between the $d$-symmetry CDW and the displacements of the atoms visualized in the $B_1$, $E_{11}$, and $E_{22}$ representations of the bond length provides a direct real-space observation of the electron-lattice interactions for the first time, indicating that the charge degree of freedom is indeed involved in the $d$CDW. 
Finally, the STM bond-length analysis offers a new paradigm for simultaneous studies of both the local lattice and electronic structures in quantum materials. 

\begin{acknowledgments}
We thank P. D. Johnson, Jiahao Yan, J. C. Davis for discussions. 
Work at Brookhaven is supported by the Office of Basic Energy Sciences, Materials Sciences and Engineering Division, U.S. Department of Energy under Contract No. DE-SC0012704. 
\end{acknowledgments}

\appendix

\begin{figure}[htbp]
\centerline{\includegraphics[scale=.25]{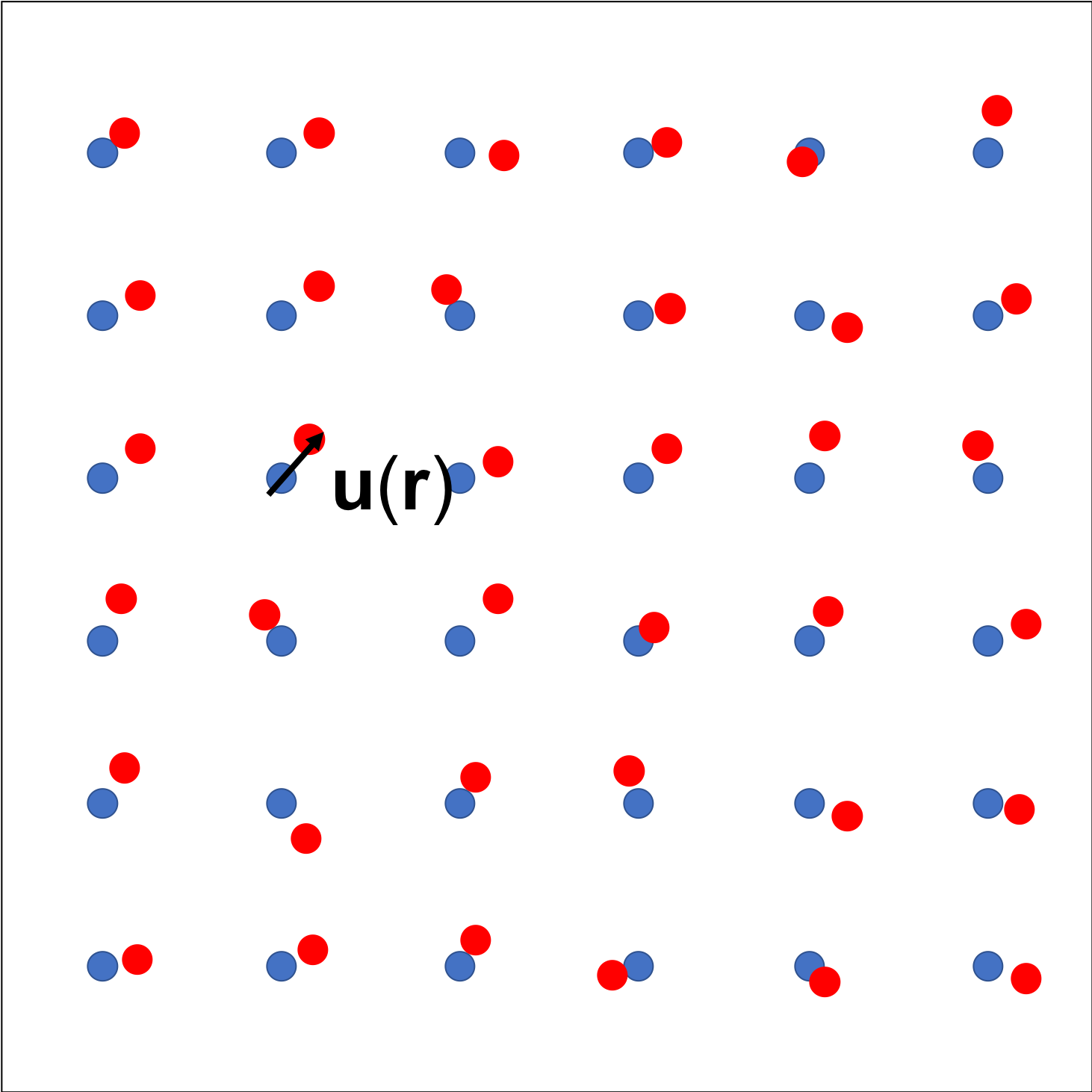}}
\caption{A schematic of the atomic displacements described by the displacement field \textbf{u}(\textbf{r}). 
The blue circles are atoms in the perfect lattice configuration. 
When there is a local lattice distortion, the lattice is displaced from the perfect lattice configuration and such a displacement is characterized by the displacement field \textbf{u}(\textbf{r}). 
}
\label{fig1s}
\end{figure}

\begin{figure}[htbp]
\centerline{\includegraphics[scale=.25]{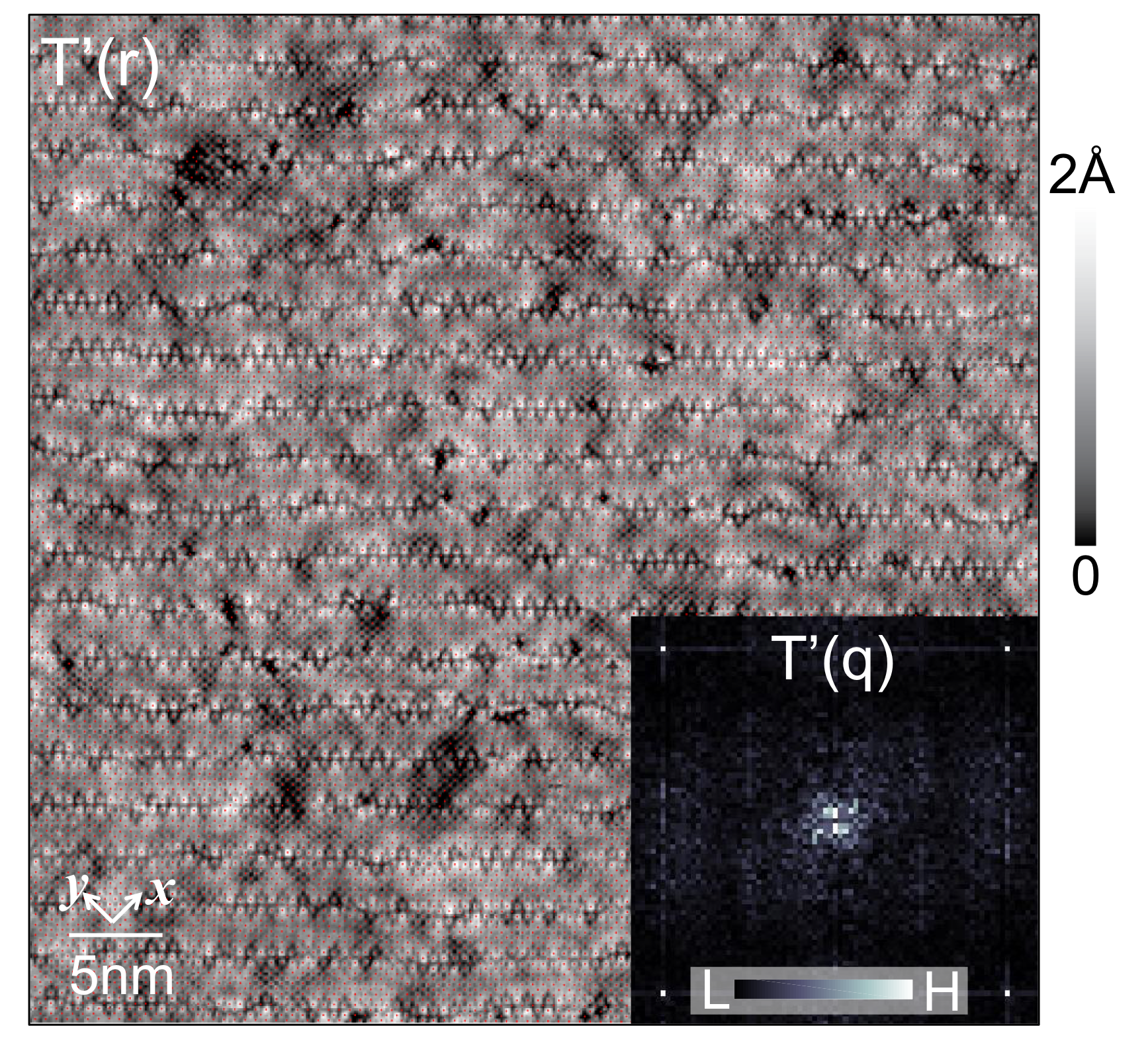}}
\caption{The same topography as shown in Fig. \ref{fig1}(a) but without the crystal structural supermodulation that is removed by the fourier filtrations at multiple supermodulation wavevectors. 
The red dots are locations of the Bi atoms. 
Inset shows the magnitude fourier transform of the topography, representing a single sharp Bragg peak.
}
\label{fig2s}
\end{figure}

\section{\label{Appendix:DetermineAtoms} Determination of the atomic sites}

\begin{figure*}[htbp]
\centerline{\includegraphics[scale=.35]{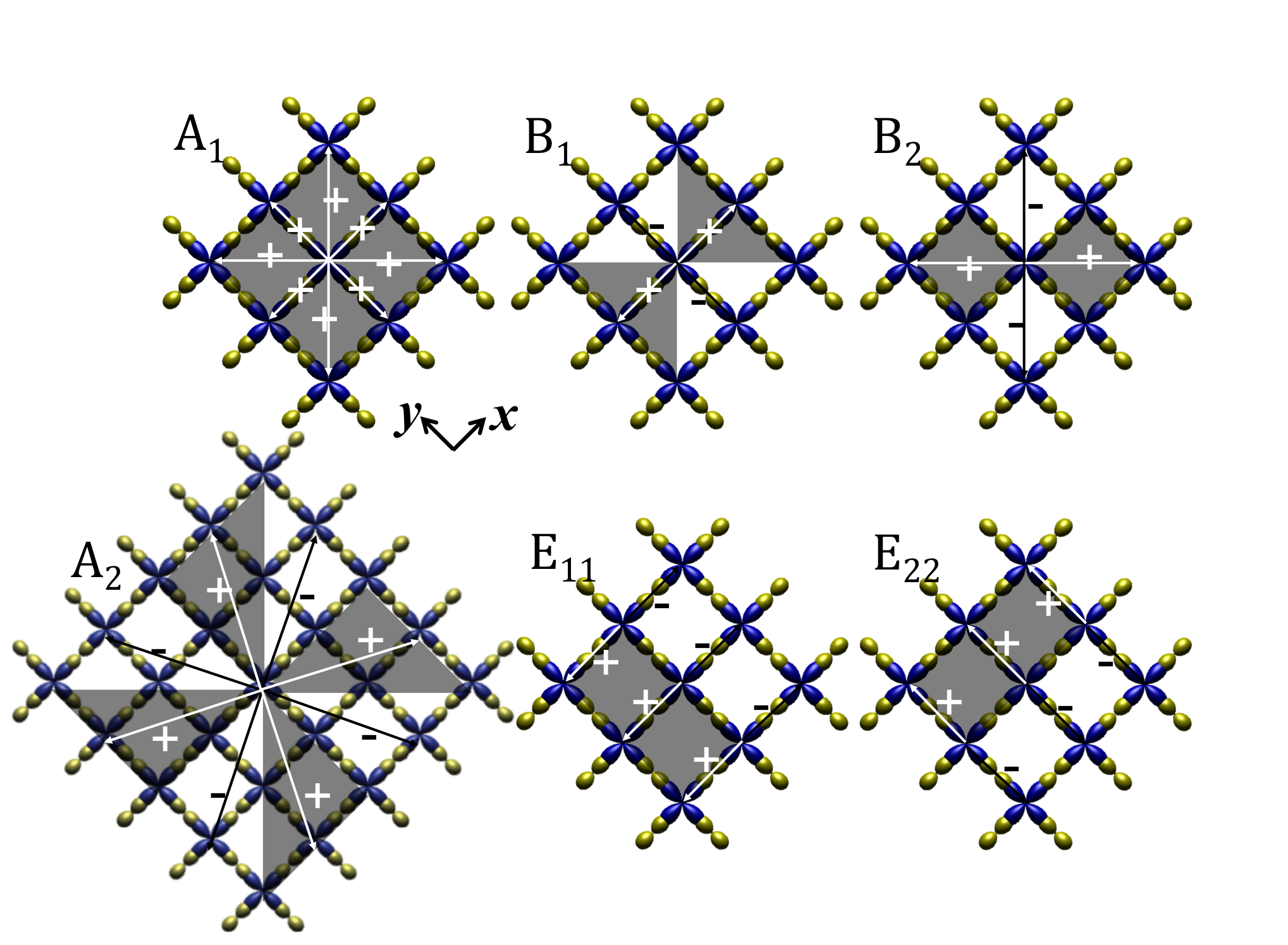}}
\caption{Procedures for calculating $S^{\Gamma}$(\textbf{r}$_i$) at the position \textbf{r}$_i$ in each representation. 
The blue (yellow) orbitals represent Cu (O$_x$ and O$_y$) sites, forming a square lattice. 
Note that Bi atoms reside ~6 \AA above the Cu sites. The symmetry dependent structure factor $s_{ij}^{\Gamma}$ in Eq (\ref{1}) in the main text is represented by the sign (and shadows) and the bond length between $i$ and adjacent site $j$ is weighted with the $s_{ij}^{\Gamma}$ and the sum is taken over the adjacent sites. 
For instance, in the case of the $B_1$ representation, the bond lengths along $x$ direction are summed and subtracted from those along $y$ direction. 
It should also be noted that $E_{11}$ and $E_{22}$ representations are modified to include contributions from the nearest neighbor sites at \textbf{r}$_i\pm$\textbf{$\delta_y$}, where \textbf{$\delta_{y}$} is a unit-cell vector along $y$ direction for $E_{11}$. 
Similarly, for $E_{22}$ to include contributions from the nearest neighbor sites at \textbf{r}$_i\pm$\textbf{$\delta_x$}, where \textbf{$\delta_x$} is a unit-cell vector along $x$ direction.
}
\label{fig3s}
\end{figure*}

\begin{table*}
\caption{Character table for the C$_{4v}$ point group}
\begin{ruledtabular}
\begin{tabular}{cccccccccc}
\multirow{2}{*}{\textrm{Basic of the representation}} &\multirow{2}{*}{\textrm{Representation}} & \multicolumn{8}{c}{Matrix represenation} \\
%\hline
 && $I$ & $C_4$ & $C_2$ & $C_4^3$ & $\sigma_x$ & $\sigma_y$ & $\sigma_d$ & $\sigma_{d'}$  \\
 \hline
 $z$ & $A_1$ & 1 & 1 & 1 & 1 & 1 & 1 & 1 & 1  \\
 \hline
 $xy(x^{2}-y^{2})$ & $A_2$ & 1 & 1 & 1 & 1 & -1 & -1 & -1 & -1 \\
 \hline
 $(x^{2}-y^{2}$)&$B_1$&1&-1&1&-1&1&1&-1&-1 \\
 \hline
 $xy$&$B_2$&1&-1&1&-1&-1&-1&1&1 \\
 \hline
 $\{x, y\}$&$E$& $\begin{pmatrix} 1 & 0 \\ 0 & 1  \end{pmatrix}$ &$\begin{pmatrix} 0 & -1 \\ 1 & 0  \end{pmatrix}$&$\begin{pmatrix} -1 & 0 \\ 0 & -1  \end{pmatrix}$&$\begin{pmatrix} 0 & 1 \\ -1 & 0  \end{pmatrix}$&$\begin{pmatrix} -1 & 0 \\ 0 & 1  \end{pmatrix}$&$\begin{pmatrix} 1 & 0 \\ 0 & -1  \end{pmatrix}$&$\begin{pmatrix} 0 & 1 \\ 1 & 0  \end{pmatrix}$&$\begin{pmatrix} 0 & -1 \\ -1 & 0  \end{pmatrix}$ \\
 \hline
\end{tabular}
\end{ruledtabular}
\label{table1}
\end{table*}

\begin{figure}[htbp]
\centerline{\includegraphics[scale=.6]{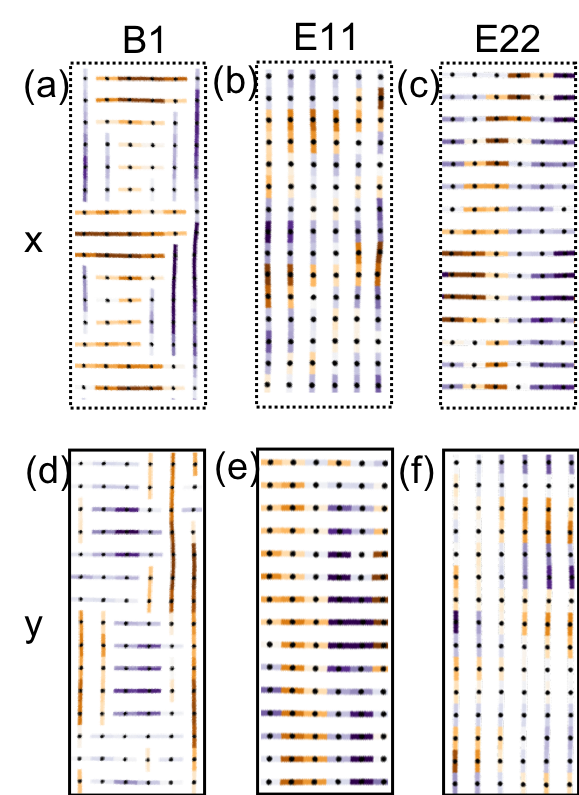}}
\caption{\textcolor{black}{ Zoomed-in $B_1$, $E_{11}$ and$E_{22}$ maps from the rectangle regions in Fig. \ref{fig1}}
}
\label{fig4s}
\end{figure}

\begin{figure}[htbp]
\centerline{\includegraphics[scale=.38]{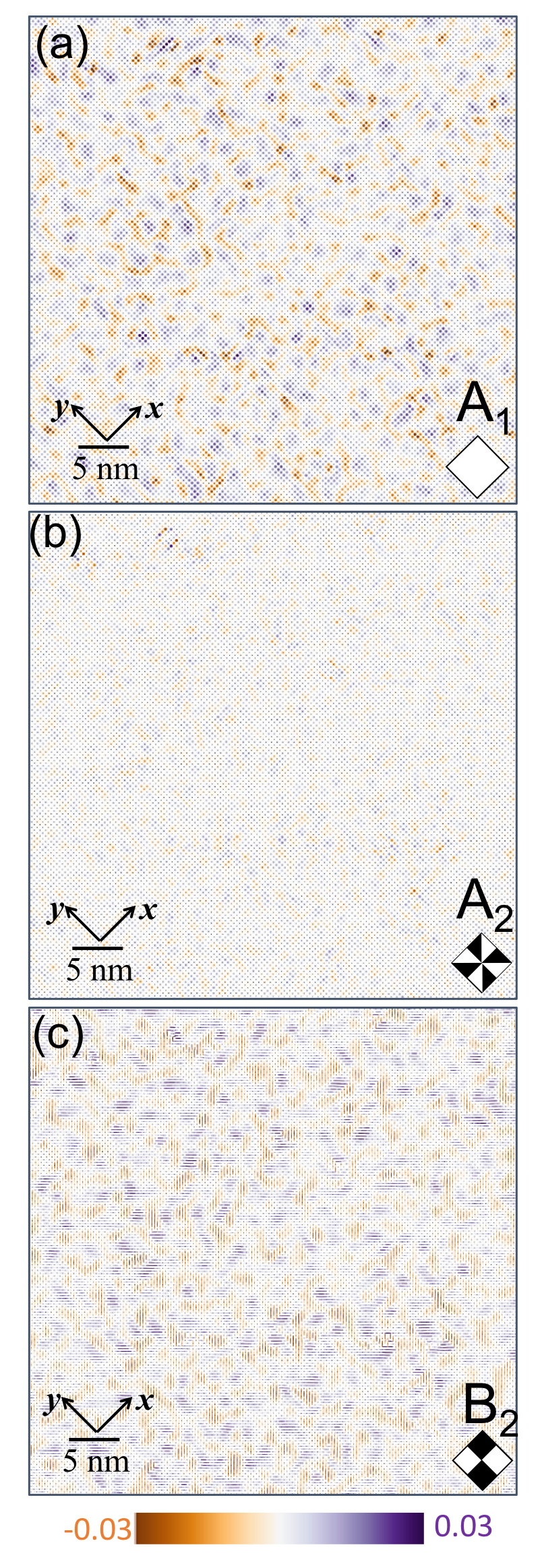}}
\caption{(a), (b), (c),$S^{A_1}$ (\textbf{r}$_i$), $S^{A_2}$ (\textbf{r}$_i$), and $S^{B_2}$ (\textbf{r}$_i$) obtained from topography shown in Fig. \ref{fig1}(a). 
}
\label{fig5s}
\end{figure}

\begin{figure*}[htbp]
\centerline{\includegraphics[scale=.38]{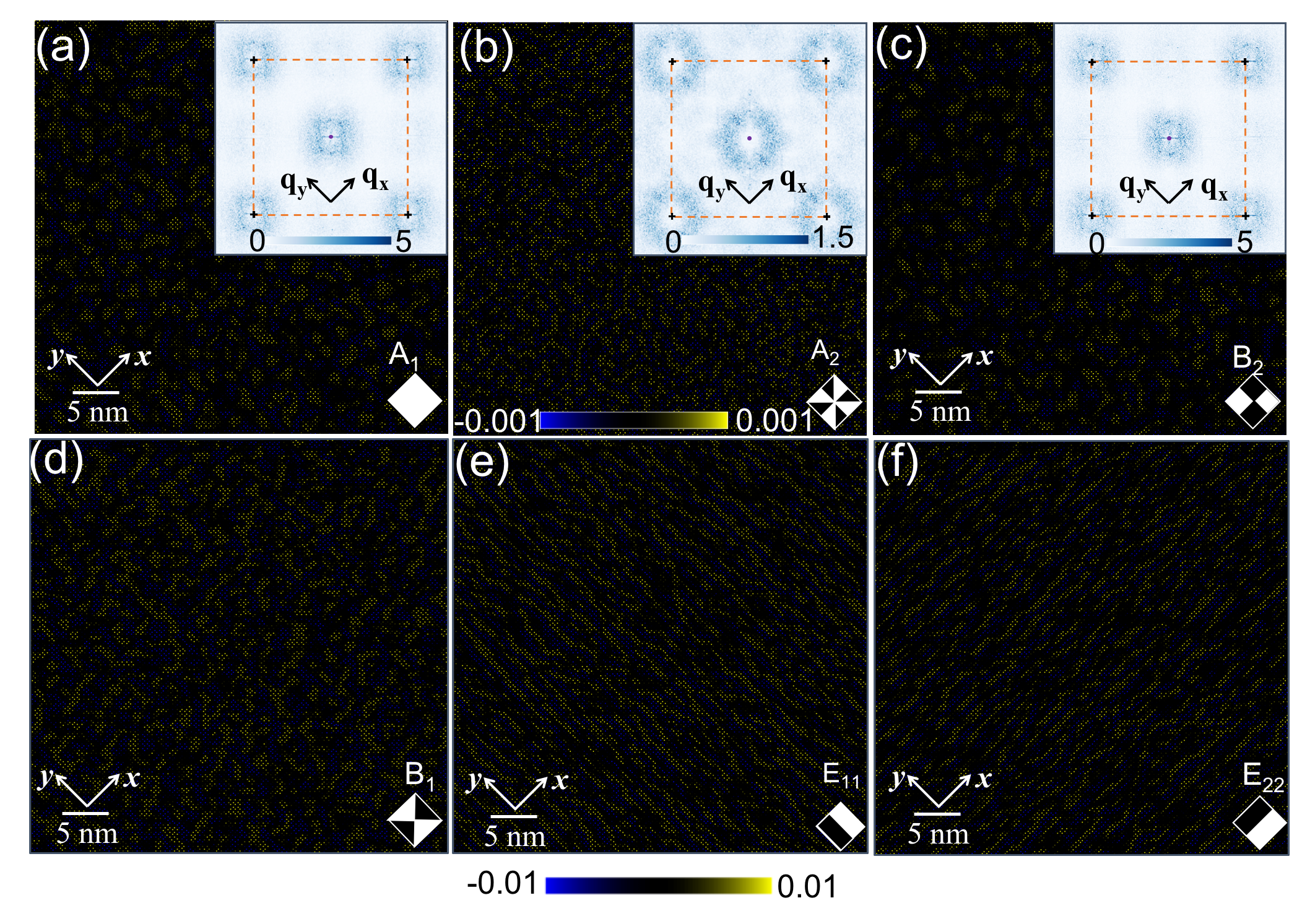}}
\caption{(a), a sublattice representation of $S^{A_1}$ (\textbf{r}$_i$) from Fig. \ref{fig4s}(a). 
Inset shows the magnitude fourier transform of (a). 
There are apparent modulations in this representation, but no well-defined fourier peaks are observed. 
(b), a sublattice representation of $S^{A_2}$ (\textbf{r}$_i$) from Fig. \ref{fig4s}(b). 
Inset shows the magnitude fourier transform of (b). (c), a sublattice representation of Fig. \ref{fig4s}(c), $B_2$(\textbf{r}). Inset shows the magnitude fourier transform of (c). 
For $A_1$, $A_2$ and $B_2$ representations, there are no well-defined peaks as shown in insets of (a), (b), (c). 
Thus, these symmetries are preserved. 
(d), (e), and (f) show $B_1$(\textbf{r}), $E_{11}$(\textbf{r}) and $E_{22}$((\textbf{r}) within the whole FOV of Fig. \ref{fig1}(a). 
}
\label{fig6s}
\end{figure*}

\begin{figure*}[htbp]
\centerline{\includegraphics[scale=.38]{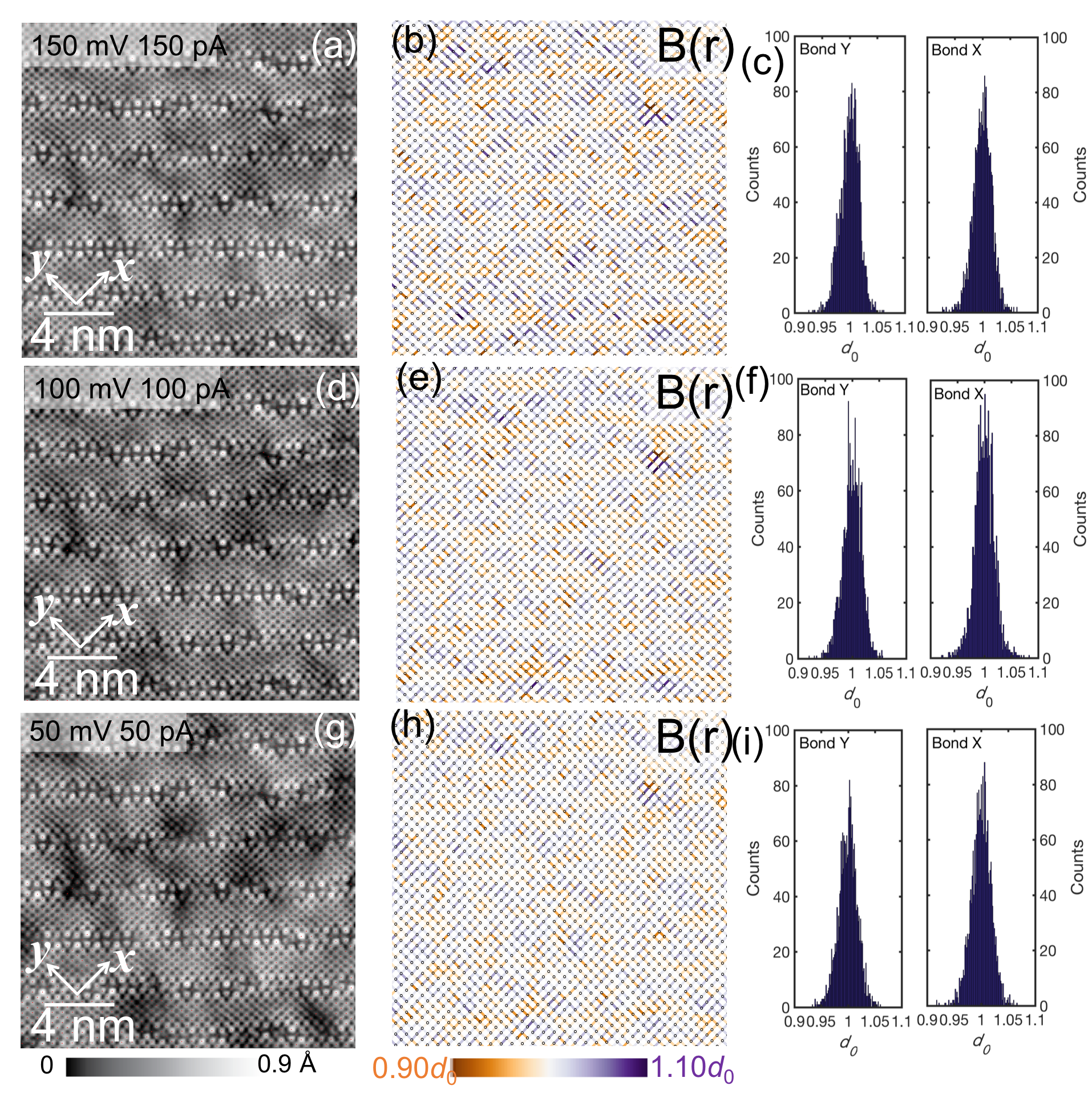}}
\caption{(a), (d), (g), topographies measured at ($I_{set}$, $V_{bias}$) = (150pA, 150 mV), (100pA, 100 mV), and (50pA, 50 mV), respectively. 
Note that the supermodulations are removed by the fourier filtrations. 
(b), (e), (h), bond maps $B(\textbf{r})$ extracted from the topographies in (a), (d), (g), respectively. 
(c), (f), (i), histograms of (b), (e), and (h), respectively, for both the $x$ and $y$ directions. 
All these maps and histograms are virtually identical, thus, a systematic error due to the junction formation in these measurements is negligible within these setup conditions. 
}
\label{fig7s}
\end{figure*}

For the purpose of the symmetry decomposition analysis, first, we apply the Affin transformations to the $T$(\textbf{r}) and the simultaneously taken electronic structure images, so that the Bragg wavevectors \textbf{Q}$_1$ and \textbf{Q}$_2$ satisfy $\vert \textbf{Q}_{1}\vert = \vert \textbf{Q}_{2} \vert$ and \textbf{Q}$_{1}\cdot \textbf{Q}_{2} =0$ after the transformation. 
Second, we apply the Lawler-Fujita algorithm [\onlinecite{LFCorrection}] to remove a STM tip drift in the data. 
In this algorithm, we obtain a displacement field \textbf{u}(\textbf{r}), which is associated with the STM tip drift, to render $T$(\textbf{r}) to be perfectly lattice periodic. 
Finally, Bi positions are determined in the topography, in which the STM tip drift is removed, but local lattice distortions remain. 
We describe these procedures step-by-step in the followings.

An original $T$(\textbf{r}), in which the STM tip drift is embedded, is given by 
%\begin{widetext}
\begin{equation}
\begin{aligned}
T(\textbf{r}) = \vert{ A_{\textbf{Q}_1}(\textbf{r})} \vert \cos(\textbf{Q}_1\cdot(\textbf{r}+\textbf{u}(\textbf{r})))\\
+\vert{ A_{\textbf{Q}_2}(\textbf{r})} \vert \cos(\textbf{Q}_2\cdot(\textbf{r}+\textbf{u}(\textbf{r})))+\dots
\label{S1}
\end{aligned}
\end{equation}
%\end{widetext}
where $\vert A_{\textbf{Q}}(\textbf{r})\vert$ is an amplitude of the complex field $A_{\textbf{Q}}(\textbf{r})$ for the wavevector \textbf{Q} and \textbf{u}(\textbf{r}) is the displacement field that distorts the $T(\textbf{r})$[\onlinecite{Chapter3}]. 
A drift-corrected topography, $T'(\textbf{r})$ is then obtained by
\begin{equation}
T'(\textbf{r})=T(\textbf{r}-\textbf{u}(\textbf{r}))
\label{S2}.
\end{equation}
Similarly, the same displacement field \textbf{u}(\textbf{r}) is also used to remove the STM-tip drift for simultaneously measured electronic structure images. 
The Lawler-Fujita algorithm utilizes a two-dimensional lock-in technique to obtain complex fields $A_{\textbf{Q}}$(\textbf{r}) for a periodic modulation at \textbf{Q}. 
Amplitude $\vert A_{\textbf{Q}}(\textbf{r})\vert$ and spatial phase shift $\Phi_{\textbf{Q}}(\textbf{r})$ are then obtained by the following equations,
\begin{equation}
A_{\textbf{Q}}(\textbf{r})=\int d\textbf{R}T(\textbf{R})e^{i\textbf{Q}\cdot\textbf{R}}e^{-\frac{(\textbf{r}-\textbf{R})^2}{2\sigma ^2}},
\label{eq:S3}
\end{equation}
\begin{equation}
|A_{\textbf{Q}}(\textbf{r})|=\sqrt{(\textrm{Re}A_{\textbf{Q}}(\textbf{r}))^2+(\textrm{Im}A_{\textbf{Q}}(\textbf{r}))^2},
\label{eq:S4}
\end{equation}
\begin{equation}
\Phi_{\textbf{Q}}(\textbf{r})=\tan^{-1}\frac{\textrm{Re}A_{\textbf{Q}}(\textbf{r})}{\textrm{Im}A_{\textbf{Q}}(\textbf{r})}.
\label{eq:S5}
\end{equation}
where \textcolor{black}{$\sigma (= L/2)$} a coarse graining length. In the case of a tetragonal lattice that is characterized by two orthogonal Bragg vectors, \textbf{Q}$_1$ and \textbf{Q}$_1$, \textbf{u}(\textbf{r}) is then simply related to the phase shift $\Phi_\textbf{Q}(\textbf{r})$ as $\textbf{Q}\cdot \textbf{u}(\textbf{r})= \Phi_{Q}(\textbf{r})$ for $\textbf{Q}=\textbf{Q}_{1}$ and $\textbf{Q}_2$, such that \textbf{u}(\textbf{r}) is obtained by
\begin{equation}
\textbf{u}(\textbf{r})=\begin{pmatrix} \textbf{Q}_1 \\\textbf{Q}_2  \end{pmatrix}^{-1}\begin{pmatrix} \Phi_{\textbf{Q}_1}(\textbf{r}) \\ \Phi_{\textbf{Q}_2}(\textbf{r})   \end{pmatrix}.
\label{S6}
\end{equation}
For the \BSCCO in this study, we use \textcolor{black}{$L$}=30 Å to calculate \textbf{u}(\textbf{r}) and the topography $T'(\textbf{r})$, in which the STM tip-drift is removed, is obtained by (\ref{S2}). 
Note that the STM tip-drift occurs at much longer length scale than 30 Å and this value is still much longer than the lattice constant of the \BSCCO. 
This means that, while extrinsic distortions in $T'(\textbf{r})$ due to the STM tip-drift are removed by (\ref{S2}), intrinsic local lattice distortions still remain in $T'(\textbf{r})$. Then, we filtered out the supermodulations.

Next, in order to locate Bi atoms, \textbf{r}$_{\textrm{Bi}}$ in $T'(\textbf{r})$, for which atomic positions are displaced from the perfect lattice configuration, we apply the Lawler-Fujita algorithm [\onlinecite{LFCorrection}] once again to $T'(\textbf{r})$ to get a displacement field \textbf{u'}(\textbf{r}) with \textcolor{black}{$L$}=6 \AA. 
Now, the shorter coarse graining length is used to capture atomic scale local lattice distortions, and \textbf{u'}(\textbf{r}) characterizes how much the atoms are displaced from the perfect lattice configuration, which is again obtained by (\ref{S6}). 
Note that the perfect lattice does not have any disorder, such that atomic corrugations are purely described by periodic functions without the phase disorder. 
Thus, a position of the Bi atoms in the distorted lattice is obtained by
\begin{equation}
\textbf{r}_{\textrm{Bi}}=\textbf{r}_{\textrm{Bi}}^\textrm{P}+\textbf{u'}(\textbf{r}_{\textrm{Bi}}^\textrm{P}),
\label{S7}
\end{equation}
where \textbf{r}$_{\textrm{Bi}}^\textrm{P}$ a vector that represents the position of the Bi atoms in the perfect lattice.
\textbf{r}$_{\textrm{Bi}}^\textrm{P}$
 is a set of vectors that are 2$\pi$ modulo in the phases of the perfect lattice,
\begin{equation}
\textbf{Q}_1 \cdot \textbf{r}_{\textrm{Bi}}^\textrm{P}=2\pi m, m=0, \pm1, \pm2, \dots, 
\label{S8}
\end{equation}
\begin{equation}
\textbf{Q}_2 \cdot \textbf{r}_{\textrm{Bi}}^\textrm{P}=2\pi n, n=0, \pm1, \pm2, \dots,
\label{S9}
\end{equation}
and then, \textbf{r}$_{\textrm{Bi}}^\textrm{P}$ is obtained by solving (\ref{S8}) and (\ref{S9}),
\begin{equation}
 \textbf{r}_{\textrm{Bi}}^\textrm{P}=2\pi \begin{pmatrix} \textbf{Q}_1 \\\textbf{Q}_2  \end{pmatrix}^{-1}\begin{pmatrix} m\\n \end{pmatrix}.
\label{S10}
\end{equation}
Finally, by plugging (\ref{S10}) into (\ref{S7}), the Bi atom positions without extrinsic distortions due to the STM tip drift but with intrinsic local lattice distortions are obtained in $T'$(\textbf{r}). 

\section{\label{Appendix:STM}SI-STM Measurements}

All the SI-STM measurements reported in this study were performed at 11K on the custom-designed SI-STM system at the OASIS complex in BNL. 
A single crystal of the nearly optimally doped \BSCCO ($T_c = 91$K, $p\sim0.17$) is cleaved by the Kapton tape at room temperature in the preparation chamber at the pressure lower than 3×10$^{-10}$ Torr. 
Prior to perform SI-STM measurements on \BSCCO, the STM tip was characterized on the single crystal Au(111) surface, which is prepared by several cycles of Ar$^{+}$ sputtering and annealing at 450$^{\circ}$C. 

\section{\label{C4v} $C_{4v}$ group}
There are eight symmetry operations \{$I$, $C_4$, $C_2$, $C_4^3$, $\sigma_{x'}$, $\sigma_{y'}$, $\sigma_d$, $\sigma_{d'}$\} and five irreducible representations \{$A_1$, $A_2$, $B_1$, $B_2$ and $E$\} in the  $C_{4v}$ point group. 
As shown in the Table \ref{table1}, a character table for the  $C_{4v}$ point group shows the corresponding matrix presentation for each irreducible representation[\onlinecite{GroupTheory}]. 
The matrix representation describes how variables are transformed by the symmetry operations for each representation. 
Figure \ref{fig3s} shows examples of how to calculate $s_{ij}^\Gamma$ for each representation. 

\section{\label{Setup} Setup condition dependence}
In order to verify that the bond map revealed in this study is not affected by the formation of the tunneling junction, which is set by the tunneling current, $I_{set}$, and $V_{bias}$ voltages, $V_{bias}$ applied between the STM tip and sample, we have systematically measured topographies at different setup conditions in the same FOV [Figs. \ref{fig6s}(a), (d), (g)]. 
At each setup conditions, bond maps are also calculated from the corresponding topographies measured. 
Figures \ref{fig6s}(a), (d), (g) show topographies measured at ($I_{set}$, $V_{bias}$) = (150pA, 150 mV), (100pA, 100 mV) and (50pA, 50 mV), respectively (1GOhm junction resistance). 
The bond maps are extracted from Figs. \ref{fig6s}(a), (d), and (g) and shown in Figs. \ref{fig6s}(b), (e) and (h), respectively. 
Similarly, corresponding histograms of the bond lengths for both the $x$ and $y$ directions for each setup conditions are shown in Figs. \ref{fig6s}(c), (f), and (i). 
The topographies, bond maps, and histograms measured at the different setup conditions virtually identical indicating that the setup effect is negligible within these conditions.

\section{\label{Cutoff} Coarse graining length dependence}
\begin{figure*}[htbp]
\centerline{\includegraphics[width=\linewidth]{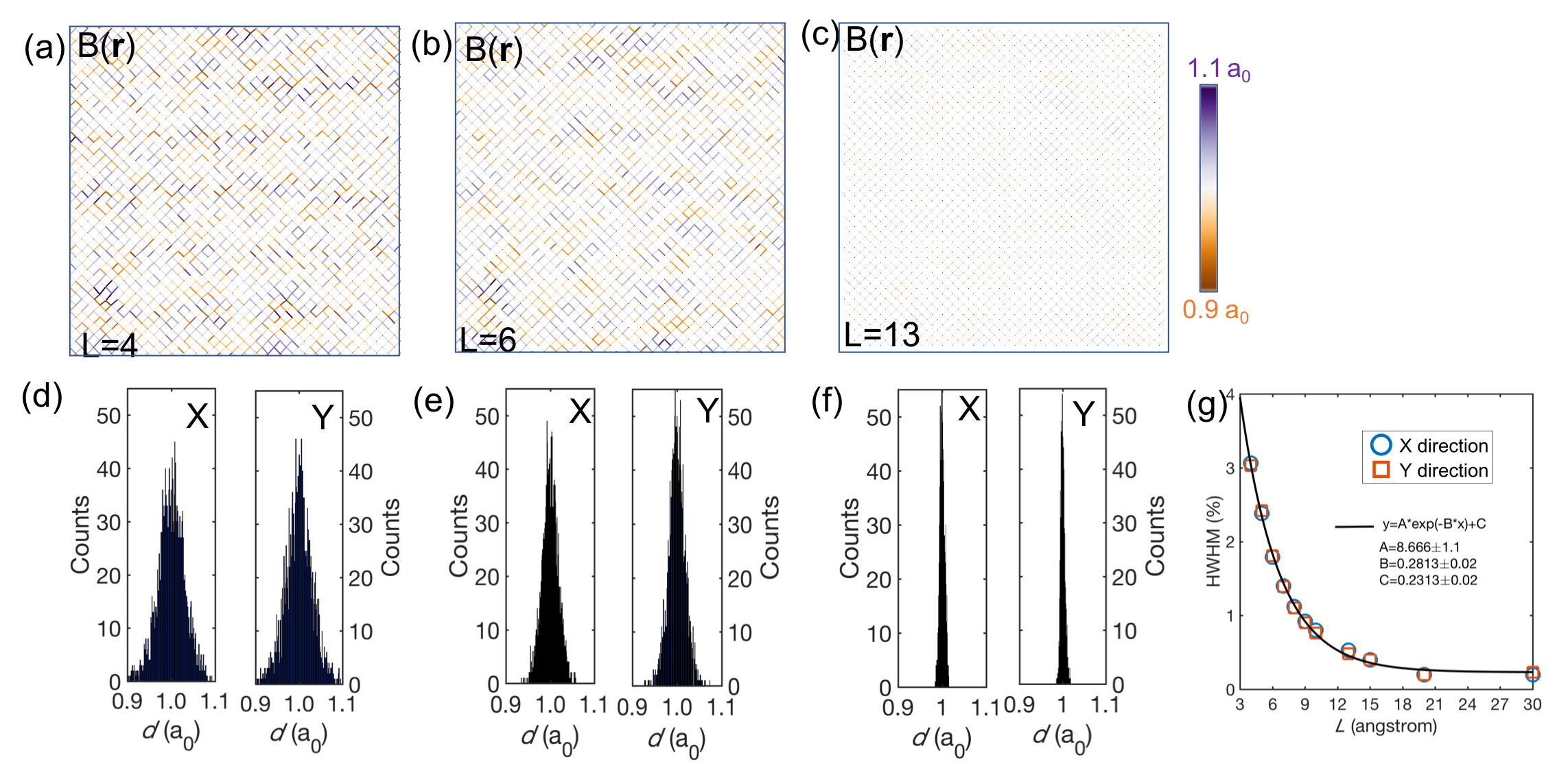}}
\caption{\textcolor{black}{(a,b,c) the bond length map obtained at different coarse-graining lengths $L$ (4, 6, 13 \AA). (d,e,f) the distributions of the bond length for both $x$ and $y$ Cu-O-Cu directions from (a,b,c). (g) the half width at half maximum of the distributions at different $L$ from 4 to 30 \AA.}}
\label{fig8s}
\end{figure*}

\textcolor{black}{In this section, we show how the local fluctuation of the bond length changes with the size of the coarse graining length $L$ in eq.(\ref{eq:S3}) and explain how we choose $L$. 
Figs. \ref{fig8s} (a,b,c) show the bond length map obtained at 4, 6, 13 \AA. 
With increasing $L$, local lattice distortions are smoothed out and washed away when $L >10$(\AA), thus the local fluctuations of the bond length become undetectable. 
For $L \leq 5$ an additional effect of the missing atoms that contributes to the bond length becomes obvious as weak horizontal contrasts since they are identified on top of the remnant supermodulations. 
Thus, $L\sim$6 (Å) is the smallest size of the filter for precisely characterizing the local lattice distortion. 
Figs. \ref{fig8s} (d), (e) and (f) show distributions of the bond length for $L$ = 4, 6, 13 \AA, respectively, from (a), (b) and (c), systematically showing a change in their height and width of the distribution. 
The half width at half maximum at each $L$ is plotted in Fig. \ref{fig8s} (g), indicating that for $L > 10$, the distribution becomes very sharp such that the large-scale fluctuations of the bond length are smoothed out and not detectable. }

\nocite{*}

\bibliographystyle{unsrt}
\bibliography{PRX_bond}% Produces the bibliography via BibTeX.

\end{document}